\documentclass[12pt]{article}
\usepackage{amssymb,amsmath,bm}
\usepackage{epsf}
\usepackage{epsfig}
\usepackage{afterpage}        
\usepackage{longtable}
\usepackage{cite}

\def\simge{\mathrel{\rlap{\raise 0.511ex \hbox{$>$}}{\lower 0.511ex \hbox{$\sim$}}}}
\def\simle{\mathrel{\rlap{\raise 0.511ex \hbox{$<$}}{\lower 0.511ex \hbox{$\sim$}}}}
\def\slash#1{\setbox0=\hbox{$#1$}\dimen0=\wd0
      \setbox1=\hbox{/} \dimen1=\wd1 \ifdim\dimen0>\dimen1
      \rlap{\hbox to \dimen0{\hfil/\hfil}} #1                        \else
      \rlap{\hbox to \dimen1{\fil$#1$\hfil}}
      /   \fi}

\newcommand{\lsim}{
\mathrel{\hbox{\rlap{\hbox{\lower4pt\hbox{$\sim$}}}\hbox{$<$}}}}

\newcommand{\gsim}{
\mathrel{\hbox{\rlap{\hbox{\lower4pt\hbox{$\sim$}}}\hbox{$>$}}}}

\newcommand{\bsi}{B_6^{(1/2)}}
\newcommand{\bei}{B_8^{(3/2)}}
\newcommand{\Lms}{\Lambda_{\overline{\rm MS}}}

\newcommand{\vcb}{|V_{cb}|}

\def\eps{\varepsilon}
\def\epe{\varepsilon'/\varepsilon}
\newcommand{\tev}{\, {\rm TeV}}
\newcommand{\gev}{\, {\rm GeV}}
\newcommand{\mev}{\, {\rm MeV}}

\newcommand{\mtb}{\overline{m}_{\rm t}}
\newcommand{\mcb}{\overline{m}_{\rm c}}

\newcommand{\mw}{M_{\rm W}}

\newcommand{\be}{\begin{equation}}
\newcommand{\ee}{\end{equation}}
\newcommand{\bea}{\begin{eqnarray}}
\newcommand{\eea}{\end{eqnarray}}
\newcommand{\nn}{\nonumber}
\newcommand{\bi}{\begin{itemize}}
\newcommand{\ei}{\end{itemize}}
\newcommand{\ord}{{\cal O}}

\newcommand{\IM}{{\rm Im}}

\def\kpn{K^+\rightarrow\pi^+\nu\bar\nu}

\def\klpn{K_{L}\rightarrow\pi^0\nu\bar\nu}

\newcommand{\newsection}[1]{\section{#1}\setcounter{equation}{0}}

\begin{document}
\begin{titlepage}
\vspace*{-0.5truecm}

\begin{flushright}
{TUM-HEP-667/07}\\
MPP-2007-43
\end{flushright}

\vfill

\begin{center}
\boldmath

{\Large\textbf{Correlations between $\epe$ and Rare $K$ Decays
\vspace{0.3truecm}\\in
    the Littlest Higgs Model with T-Parity}}

\unboldmath
\end{center}

\vspace{0.4truecm}

\begin{center}
{\bf Monika Blanke$^{a,b}$, Andrzej J.~Buras$^a$,  Stefan Recksiegel$^a$,\\
Cecilia  Tarantino$^a$ and Selma Uhlig$^a$}
\vspace{0.4truecm}

 $^a${\sl Physik Department, Technische Universit\"at M\"unchen,
D-85748 Garching, Germany}

 {\sl $^b$Max-Planck-Institut f{\"u}r Physik (Werner-Heisenberg-Institut), \\
D-80805 M{\"u}nchen, Germany}
\end{center}
\vspace{0.6cm}
\begin{abstract}
\vspace{0.2cm}
\noindent 
We calculate the CP-violating ratio $\epe$ in the Littlest Higgs model with
T-parity (LHT) and investigate its correlations with the branching ratios for
$\klpn$, $K_L\to\pi^0\ell^+\ell^-$ and $\kpn$. The resulting correlations are
rather strong in the case of $K_L$ decays, but less pronounced in the case of
$\kpn$. Unfortunately, they are subject to large hadronic uncertainties present
in $\epe$, whose theoretical prediction in the Standard Model (SM) is reviewed
  and updated here. With the matrix elements of $\mathcal{Q}_6$ (gluon penguin) and
$\mathcal{Q}_8$ (electroweak penguin) evaluated in the large-$N$ limit and
$m_s^{\overline{\rm MS}}(2\gev) = 100\mev$ from lattice QCD, $(\epe)_\text{SM}$
turns out to be close to the data so that significant departures of
$Br(\klpn)$ and $Br(K_L\to\pi^0\ell^+\ell^-)$ from the SM expectations are unlikely, while $Br(\kpn)$ can be enhanced even by a factor 5. On the other hand, modest departures of the relevant hadronic matrix elements from their large-$N$ values allow for a consistent description of $\epe$ within the LHT model accompanied by large enhancements of $Br(\klpn)$ and $Br(K_L\to\pi^0\ell^+\ell^-)$, but only modest enhancements of $Br(\kpn)$.
\end{abstract}

\vfill
\end{titlepage}

\thispagestyle{empty}

\begin{center}
{\Large\bf Note added}
\end{center}

\noindent
An additional contribution to the $Z$ penguin in the Littlest Higgs model with T-parity has been pointed out in \cite{Goto:2008fj,delAguila:2008zu}, which has been overlooked in the present analysis. This contribution leads to the cancellation of the left-over quadratic divergence in the calculation of some rare decay amplitudes. Instead of presenting separate errata to the present work and our papers \cite{Blanke:2006eb,Blanke:2007db,Blanke:2007ee,Blanke:2008ac} partially affected by this omission, we have presented a corrected and updated analysis of flavour changing neutral current processes in the Littlest Higgs model with T-parity in \cite{Blanke:2009am}.

\newpage

\setcounter{page}{1}
\pagenumbering{arabic}

\newsection{Introduction}
\label{sec:intro}
Flavour Changing Neutral Current (FCNC) processes provide a powerful tool for
testing the Standard Model (SM) and its extensions.
Of particular interest are the four rare kaon decays $K_L \to \pi^0 \nu \bar
\nu$, $K^+ \to \pi^+ \nu \bar \nu$, $K_L \to \pi^0 e^+ e^-$ and $K_L \to \pi^0
\mu^+ \mu^-$.
Their branching ratios are strongly suppressed within the SM and consequently
can be largely modified by New Physics (NP) contributions.

Extensive analyses of these decays in the MSSM {\cite{Isidori:2006qy}}, the Littlest Higgs model
with T-parity (LHT) \cite{Blanke:2006eb}, general models with enhanced $Z$-penguin 
contributions \cite{BFRS} and $Z'$-models \cite{Zprime} have shown that in the presence 
of new sources of flavour and CP-violation beyond those present in the Minimal 
Flavour Violation (MFV) framework \cite{mfv1,mfv2}, enhancements of 
$Br(K_L \to \pi^0 \nu \bar \nu)$ by an order of magnitude and of the other 
branching ratios by up to a factor $5$ are still possible.

On the other hand, as pointed out in \cite{BS98} and analyzed in more detail
within the MSSM in \cite{BCIRS99}, the enhancements of the rare decay branching ratios
in question could be bounded in principle by the value of
$\epe$ that measures the ratio of the direct and indirect
CP-violating contributions to $K_L \to \pi \pi$.
The reason is very simple.
The electroweak penguin and box diagrams that enter the evaluation of the rare decay branching ratios in question have also considerable impact on the ratio
$\epe$ so that, in a given model, specific correlations
between $\epe$ and the branching ratios for rare $K$
decays exist.

Unfortunately, whereas the branching ratios of $K \to \pi \nu \bar \nu$ decays
are theoretically very clean {\cite{Buras:2004uu}} and those of $K_L \to \pi^0 \ell^+ \ell^-$ are
subject to only moderate theoretical uncertainties {\cite{Buchalla:2003sj}}, this is not the case for
the ratio $\epe$ that is affected by large hadronic uncertainties.

Indeed, whereas the Wilson coefficients of the local operators entering the
evaluation of $\epe$ are known \cite{Buras:1993dy,Ciuchini:1992tj,Buras:1991jm,Buras:1992tc,Buras:1992zv,Ciuchini:1993vr} at the NLO level in
QCD and QED renormalization group improved perturbation theory, the
hadronic matrix elements of these operators are still only poorly
known.\footnote{Latest reviews can be found in {\cite{BJ03,Bijnens:2003hk,Pich:2004ee,Dawson:2005za,Lee:2006cm}}.}
Therefore the predictions for $\epe$ in the SM and its
extensions have very large theoretical uncertainties.

In spite of this unsatisfactory situation and in view of future improvements
in the evaluation of the relevant hadronic matrix elements by lattice QCD or large-$N$ methods, we think that it is important to analyze the correlations between 
$\epe$ and rare kaon decays in specific extensions of the
SM, where large enhancements of the rare decay branching ratios have been
found.
Certainly, the result of such an exercise will sensitively depend on the
values of the hadronic parameters present in $\epe$, but
the mere fact that such correlations exist will hopefully motivate further
non-perturbative studies.

The main goal of the present paper is the calculation of
$\epe$ within the LHT model \cite{ACKN,CL} and the investigation of its
correlations with the four rare kaon decays in question, for a given set of
the non-perturbative parameters entering  $\epe$.
To this end we will apply a useful parameterization of $\epe$
proposed in \cite{BJ03} that automatically takes into account all renormalization
group effects from scales below $m_t$ and expresses the hadronic uncertainties
in terms of the two parameters $R_6$ and $R_8$ corresponding to the dominant
QCD and electroweak penguin operators, respectively.

In \cite{Blanke:2006eb} very sharp and theoretically clean correlations between the decays
 $K_L \to \pi^0 \nu \bar \nu$, $K^+ \to \pi^+ \nu \bar \nu$ and $K_L \to \pi^0
 \ell^+ \ell^-$ have been found in the LHT model,
subject mainly to a discrete ambiguity present in the correlation between $\klpn$ and $\kpn$.
It is therefore sufficient to establish the correlations between
$\epe$ and  $K_L \to \pi^0 \nu \bar \nu$ and between $\epe$ and $\kpn$ in order to get
 an idea about all correlations.

Our paper is organized as follows. In Section \ref{sec:SM} we briefly review
the status of $\epe$ in the SM, investigate the relevant theoretical and
parametric uncertainties and provide a numerical update of \cite{BJ03}.
In Section~\ref{sec:epsprime} we present the basic formulae for
$\epe$ in a generic model with new complex phases but no
new operators relative to the SM, in terms of the short distance functions
$X,Y,Z$ and $E$ that contain both SM and NP contributions.
It turns out that in the LHT model the functions $X,Y$ and $Z$ can directly be
obtained from our previous analysis \cite{Blanke:2006eb}.
The function $E$ that plays a subdominant role in $\epe$,
is calculated for completeness here for the first time in the LHT model.
In Section~\ref{sec:numerics} we evaluate $\epe$ in the
LHT model scanning over its parameters and for various
values of $R_6$ and $R_8$.
The main results of this section are the correlations between
$\epe$ and  $Br(K_L \to \pi^0 \nu \bar \nu)$ and between $\epe$ and $Br(\kpn)$, that
illustrate the very important role of $\epe$ in bounding
the enhancements of rare $K$
decay branching ratios provided the non-perturbative parameters $R_6$ and
$R_8$ are accurately known.  
We conclude in Section~\ref{sec:conclusions}.

\boldmath
\newsection{$\epe$ in the SM}
\unboldmath
\label{sec:SM}

\subsection{Basic Formula}
Before analyzing $\epe$ within the LHT model, it will be instructive to have a brief look at this ratio within the SM and investigate the relevant theoretical and parametric uncertainties that have to be taken into account also in the case of the LHT model. This will also allow us to update the analysis of \cite{BJ03}.

The formula for $\epe$ of \cite{BJ03} is given in the SM as follows:
\be\label{epeSM}
\frac{\eps'}{\eps} = \IM (\lambda_t) \cdot \Big[ P_0 + P_E E_0(x_t) + P_X X_0(x_t) + P_Y Y_0(x_t) + P_Z Z_0(x_t) \Big]
\ee
with $\lambda_t = V_{ts}^* V_{td}^{}$ and $x_t={m_t^2}/{M_{W}^2}$.
The short distance physics is described by the loop functions $E_0(x_t)$, $X_0(x_t)$, $Y_0(x_t)$ and $Z_0(x_t)$, for which explicit expressions can be found in \cite{Buchalla:1995vs}.
On the other hand, the $P_i$ encode information about the physics at scales 
$\mu \le\ord(m_t, M_W)$, and are given in terms of 
the hadronic parameters
\be\label{RS}
R_6\equiv \bsi\left[ \frac{121\mev}{m_s(m_c)+m_d(m_c)} \right]^2,\quad
R_8\equiv \bei\left[ \frac{121\mev}{m_s(m_c)+m_d(m_c)} \right]^2
\ee 
as follows:
\begin{equation}
P_i = r_i^{(0)} + 
r_i^{(6)} R_6 + r_i^{(8)} R_8 .
\label{eq:pbePi}
\end{equation}
The coefficients $r_i^{(0)}$, $r_i^{(6)}$ and $r_i^{(8)}$ enclose
information on the Wilson-coefficient functions of the $\Delta S=1$ weak
effective Hamiltonian at the next-to-leading order
\cite{Buchalla:1995vs}.
Their numerical values for
different choices of $\Lms^{(4)}$ at $\mu=m_c$ in the NDR renormalization
scheme can be found in \cite{BJ03}.
The numerical values 
of the $P_i$ are sensitive functions of $R_6$ and $R_8$, as well as of
$\Lms^{(4)}$ or equivalently $\alpha_s(M_Z)$.
The values $\Lms^{(4)}=310,340,370\mev$ considered in \cite{BJ03} and by us correspond
  to the three-loop values $\alpha_s(M_Z)=0.119, 0.121, 0.123$,
  respectively. The two-loop formula for the strong coupling constant,
  instead, provides $\alpha_s(M_Z)=0.117, 0.119, 0.121$. Although three-loop
  values are quoted by the PDG \cite{PDG,Bethke:2006ac}, we use in the present analysis the two-loop values, as the Wilson coefficients entering
  $\epe$ are known at the NLO only.

\boldmath
\subsection{Status of $\bsi$ and $\bei$ from Lattice QCD}
\unboldmath
The hadronic parameters $\bsi$ and $\bei$  represent 
the matrix elements of the dominant QCD penguin operator $\mathcal{Q}_6$ and the 
dominant EW penguin operator $\mathcal{Q}_8$, respectively.
They are the main source of uncertainty in the determination of $\epe$ and,
hence, calculating $\langle \pi
\pi |\mathcal{Q}_{6,8} |K\rangle$ reliably represents a theoretical challenge
{for the non-perturbative methods like lattice QCD and large-$N$.
The large-$N$ approach will be referred to below while we focus
here on the status of lattice studies relevant for $\epe$.}
The lattice calculation of the $\mathcal{Q}_6$ matrix element is particularly delicate.
Golterman and Pallante \cite{Golterman:2001qj}, indeed, have pointed out that there is a serious
ambiguity in the lattice version of left-right QCD penguin operators, like
$\mathcal{Q}_6$, because the flavour group in (partially) quenched QCD is not $SU(3)$
but $SU(3+N_f|3)$ where $N_f$ is the number of sea quark flavours.
It {turns out} that the ambiguity in $\mathcal{Q}_6$ has such a large effect on $\epe$
that it can even flip its sign in quenched QCD \cite{Bhattacharya:2004qu,Bhattacharya:2001uw,Aubin:2006vt}.
Moreover, the same problem affects the left-left QCD penguin operator $\mathcal{Q}_4$
with a sub-leading effect in $\epe$ \cite{Golterman:2006ed}.
On the other hand, the lattice calculation of the $\mathcal{Q}_8$ matrix element is more
reliable, although challenging as well and still affected by an uncertainty of
$10 \div 20$\%.
Two independent approaches have been used.
In the indirect approach, one calculates the hadronic matrix elements of $K
\to \pi$ and $K \to 0$ and reconstructs $K \to \pi \pi$ amplitudes using chiral
perturbation theory. This method, relatively easy and computationally cheap,
has been widely used \cite{Bhattacharya:2004qu,Blum:2001xb,Noaki:2001un}, but it only works in leading order chiral
perturbation theory.
In the direct approach, instead, one calculates directly the $K \to \pi \pi$ matrix
elements with the final state pions carrying a physical momentum.
The difficulty of this method is represented by the Maiani-Testa ``no-go
theorem'' \cite{Maiani:1990ca}: one can not obtain $K \to \pi(\vec p) \pi(-\vec p)$ but only $K \to
\pi(\vec 0) \pi(\vec 0)$ on the lattice, where $| \pi(\vec 0) \pi(\vec
0)\rangle$ is the ground state of two pions with periodic boundary condition
in the spatial direction.
Various methods have been proposed to get around the Maiani-Testa theorem. 
Luscher and Wolff \cite{Luscher:1990ck} proposed a
diagonalization method, based on a
computationally expensive calculation of correlators with non-zero
pion momentum.
Another possibility consists in modifying the boundary condition for the pions
\cite{Kim:2005gk}, thus providing a finite momentum to the ground state of $\pi^\pm$.
A different approach was elaborated by Lellouch and Luscher \cite{Lellouch:2000pv}, based on
an excited state fit to extract the $|\pi(\vec p) \pi(-\vec p)\rangle$ state
that appears in a finite volume where the spectrum of two-particle states is
discrete, and on a formula for connecting the decay measured
in a finite volume to the infinite volume result, in the center of mass (CM) frame.
This technique, however, is challenging due to the need to extract the excited
state.
An alternative and promising method is to work with a kinematic setup for which the final
state of interest is also the lowest energy state. This has been done in
\cite{Boucaud:2004aa,Boucaud:2001tx,Lin:2002nq}
by working in the moving (LAB) frame, i.e. calculating $\langle \pi(\vec P)
\pi (\vec 0) |\mathcal{Q}_8 |K(\vec P)\rangle$  and then converting the result
from the finite to the infinite volume, using the
Lellouch-Luscher formula \cite{Lellouch:2000pv}.
An important theoretical advance of the last years is the derivation of a
relationship similar to the Lellouch-Luscher formula but valid in the LAB frame
\cite{Christ:2005gi,Kim:2005gf} that may improve the accuracy of the LAB-frame method, as shown
in a preliminary calculation with domain wall fermions \cite{Yamazaki:2006ce}.

\subsection{{Comparison between SM Prediction and Experimental Data}}
On the experimental side, the world average based on the latest 
results from NA48 \cite{NA48} and KTeV
\cite{KTeV} and previous results from NA31 \cite{Barr:1993rx} and E731
\cite{Gibbons:1993zq} reads 
\begin{equation}\label{eps}
\epe=(16.7\pm 1.6) \cdot 10^{-4}.
\end{equation}
While several analyses made in recent years within the SM found results 
that are compatible with (\ref{eps}), it is fair to say that the 
large hadronic uncertainties  in the coefficients $P_i$ still allow for 
sizeable NP contributions. The relevant list of references can be
found in \cite{Pich:2004ee,Bijnens:2003hk,Dawson:2005za,Lee:2006cm,BJ03}.

In \cite{BJ03} an agreement of the SM with \eqref{eps} has been found for
$(R_6,R_8)=(1.2,1.0)$ and $\Lms^{(4)}=(340\pm30)\mev$. Meanwhile the value of
$m_t$ decreased and the value of $\IM(\lambda_t)$ increased. Consequently for
$R_6=R_8=1.0$, corresponding to the large-$N$ approach of \cite{Bardeen:1986vp} with
$B_6^{(1/2)}=B_8^{(3/2)}=1$,   and $m_s^{\overline{\rm MS}}(2\gev)=100\mev$
from lattice QCD  \cite{PDG,mstrange},\footnote{Similar results are found
  from QCD sum rules \cite{Jamin:2006tj}.} acceptable agreement with \eqref{eps} can be
obtained, provided $\Lms^{(4)}>340\mev$. Indeed in this case we find for $\Lms^{(4)}=340\mev$ 
\be\label{PI1}
P_0=15.962\,,\quad P_X=0.597\,,\quad P_Y=0.519\,,\quad P_Z=-12.416\,,\quad P_E=-1.226\,,
\ee
and choosing $\IM(\lambda_t)=1.38 \cdot 10^{-4}$, obtained by the UTfit collaboration \cite{UTFIT}, the result
\be\label{eq:SM-UTfit-0.119}
(\epe)_\text{SM}=12.3\cdot 10^{-4}
\ee
which is a bit lower than the value in \eqref{eps}. For $\Lms^{(4)}=370\mev$ we find, on the other hand,
\be\label{eq:SM-UTfit-0.121}
(\epe)_\text{SM}=13.5\cdot 10^{-4}\,,
\ee
within $2\sigma$ from the central value in \eqref{eps}.
A slight decrease of the $m_s^{\overline{\rm MS}}(2\gev)$ value (see Table
\ref{tab:Rms}) would result in an improved agreement with the data.

We would like to emphasize, then,  that with $\IM(\lambda_t)= 1.69 \cdot
10^{-4}$, obtained from the tree level determination of the CKM parameters, the values in \eqref{eq:SM-UTfit-0.119} and \eqref{eq:SM-UTfit-0.121} increase to 
\be\label{eq:SM-tree-0.119}
(\epe)_\text{SM}=15.3\cdot10^{-4}\,,
\ee
and
\be\label{eq:SM-tree-0.121}
(\epe)_\text{SM}=16.7\cdot10^{-4}\,,
\ee
so that even for  $\Lms^{(4)}=340\mev$ and $m_s^{\overline{\rm
    MS}}(2\gev)=100\mev$ a good agreement with the data can be obtained.

\begin{figure}
\begin{center}
\epsfig{file=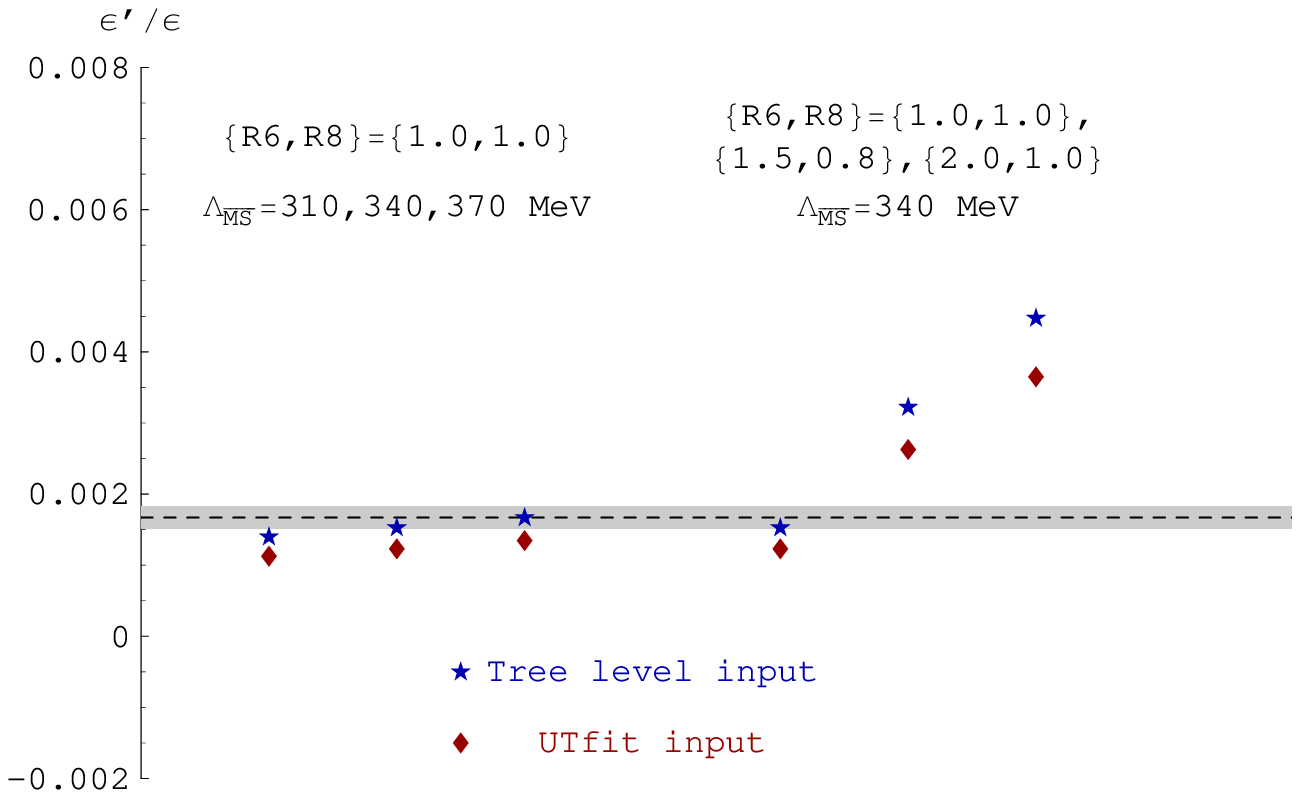,scale=0.8}
\end{center}
\caption{\label{fig:errors}\it $(\epe)_\text{SM}$ for three different choices
  of $(R_6,R_8)=(1.0,1.0),(1.5,0.8),(2.0,1.0)$ and different values of
  $\Lms^{(4)}=310,340,370\mev$. The values obtained with the UTfit value for
  $\IM(\lambda_t)^\text{UTfit}=1.38 \cdot 10^{-4}$ are marked with red diamonds, while
  those with the tree level value $\IM(\lambda_t)^\text{tree}= 1.69 \cdot
  10^{-4}$ are marked with blue stars. The shaded area represents the experimental result in \eqref{eps}.}
\end{figure}

As a preparation for the analysis of $\epe$ in the LHT model we show in Fig.~\ref{fig:errors} the values of  $(\epe)_\text{SM}$ for three different choices of $(R_6,R_8)=(1.0,1.0),(1.5,0.8),(2.0,1.0)$, different values of $\Lms^{(4)}$ and the two values for  $\IM(\lambda_t)$ considered above.

The main messages from Fig.~\ref{fig:errors} and  \eqref{eq:SM-UTfit-0.119}-\eqref{eq:SM-tree-0.121} are:
\bi
\item
$(\epe)_\text{SM}$ has a visible dependence on the values chosen for $\IM(\lambda_t)$ and for $\Lms^{(4)}$, but these dependences amount only to about $10\div20$\%, which is comparable to the experimental error in \eqref{eps}.
\item
 $(\epe)_\text{SM}$ depends very strongly on the values of $R_6$ and $R_8$,
 and the choices $(1.5,0.8)$ and $(2.0,1.0)$ give values for
 $(\epe)_\text{SM}$ that clearly are in disagreement with the data for the
 full range of $\Lms^{(4)}$ and $\IM(\lambda_t)$ considered by us.
For instance for $\Lms^{(4)}=340 \mev$ and the UTfit value of
 $\IM(\lambda_t)$ one finds $(\epe)_\text{SM}=26.3\cdot10^{-4}$ and $(\epe)_\text{SM}=36.5\cdot10^{-4}$ for
 $(R_6,R_8)=(1.5,0.8)$ and $(R_6,R_8)=(2.0,1.0)$, respectively.
\item
Significant although smaller departures of $(R_6,R_8)$ from $(1.0,1.0)$ 
  and therefore of $\epe$ from the data could also occur
 for $B_6^{(1/2)}=B_8^{(3/2)}=1$, as obtained from the large-$N$ approach of \cite{Bardeen:1986vp},  and
 values of the strange quark mass deviating from
 $m_s^{\overline{\rm MS}}(2\gev)=100\mev$ by the present $10 \div 20$\%
 lattice uncertainty (see
Table \ref{tab:Rms}).
\ei

\begin{table}[ht]
{\renewcommand{\arraystretch}{1.1}\setlength{\arraycolsep}{2pt}
\begin{center}
\begin{tabular}{|c || c|c|c|c|c|}
\hline
$m_s^{\overline{\rm MS}}(2\gev)$ & $80\mev$ & $90\mev$ & $100\mev$ & $110\mev$ & $120\mev$\\
\hline
$(R_6,R_8)$ & $(1.5,1.5)$ & $(1.2,1.2)$ & $(1.0,1.0)$  & $(0.8,0.8)$ & $(0.7,0.7)$\\ \hline
\end{tabular}
\end{center}}
\caption{\label{tab:Rms}\it  Choices for the strange quark mass within
  present lattice uncertainties and corresponding
 values for the hadronic parameters $(R_6,R_8)$. The small down quark mass has a
 minor impact and its value is fixed to $m_d^{\overline{\rm MS}}(2\gev)=5
 \mev$ \cite{PDG}. The variation of $\Lms^{(4)}$ entering the quark mass
 running represents a small effect as well and its value is fixed to $\Lms^{(4)}=340 \mev$.}
\end{table}

As reviewed in \cite{BJ03}, $R_8=1.0\pm0.2$ is obtained in various
approaches. Unfortunately the value of $R_6$ is very uncertain. For instance
in the large-$N$ approach of \cite{B6A,B6B} values for $R_6$ significantly higher than $1$ have been
found. In particular \cite{B6A} reports $R_6=2.2\pm0.4$ and
$R_8=1.1\pm0.3$. On the other hand, while the lattice values of $R_8$ are
compatible with $1$ \cite{Boucaud:2004aa,Blum:2001xb}, they are
lower than unity for $R_6$ \cite{Bhattacharya:2004qu,Bhattacharya:2001uw}.

\boldmath
\newsection{$\epe$ in the LHT Model}
\unboldmath
\label{sec:epsprime}
The LHT model \cite{ACKN,CL} belongs to the class of Little Higgs models
\cite{ACG}, where the little hierarchy problem is solved by a naturally light
Higgs, identified with a Nambu-Goldstone 
boson of a spontaneously broken global
symmetry.
In the LHT model 
the global group $SU(5)$ is spontaneously broken
into $SO(5)$ at the scale $f \approx \mathcal{O}(1 \tev)$ and
the electroweak sector of the SM is embedded in an $SU(5)/SO(5)$ non-linear
sigma model. 
Gauge and Yukawa Higgs interactions are introduced by gauging the subgroup of
$SU(5)$: $[SU(2) \times U(1)]_1 \times [SU(2) \times U(1)]_2$, such that the
so-called collective symmetry breaking prevents the Higgs from becoming
massive when the couplings of one of the two gauge factors vanish.
A discrete symmetry called T-parity \cite{CL} is then introduced, in order to reconcile
the model with electroweak precision tests.
It restores the custodial $SU(2)$ symmetry and, therefore, the compatibility with
electroweak precision data is obtained already for quite small values of the
NP scale, $f \ge 500 \gev$ \cite{HMNP,AMOO}.
Another important consequence is that particle fields are T-even or T-odd
under T-parity.
The particles belonging to the T-even sector are the SM particles and a
heavy top $T_+$, while the T-odd sector consists of heavy gauge bosons
$W_H^\pm,Z_H,A_H$, a
scalar triplet $\Phi$, 
an odd heavy top $T_-$ and the so-called mirror fermions \cite{Low:2004xc}, i.e.,
fermions corresponding to the SM ones but with opposite T-parity and $\mathcal{O}(1 \tev)$ mass.
Mirror fermions are characterized by new flavour {and CP-violating} interactions with SM fermions
and heavy gauge bosons, thus allowing significant 
effects in flavour observables \cite{Blanke:2006eb,BBPTUW,HLP,Blanke:2007ee,Indian,Blanke:2007db} without new operators in addition to the SM
ones.

The formula for the CP-violating ratio $\epe$ of \cite{BJ03} in a generic
model with new complex phases but no new operators, like the LHT model,
generalizes as follows: 
\be 
\frac{\varepsilon'}{\varepsilon}= \frac{\IM(\lambda_t)}{\sin(\beta-\beta_s)}\, \tilde F_{\varepsilon'}(v),
\label{epeth}
\ee
with $\lambda_t =V_{ts}^* V_{td}^{}$, $\beta_s = -1.3^\circ$ and
\bea
\tilde F_{\varepsilon'}(v) &=&P_0 \sin(\beta-\beta_s) + P_E \,
|E_K|\sin\beta_E^K\nn\\
&& + P_X \, |X_K|\sin\beta_X^K + 
P_Y \, |Y_K|\sin\beta_Y^K
 + P_Z \, |Z_K|\sin\beta_Z^K\,,
\label{FE}
\eea
where $\beta$ is the angle in the unitarity triangle to be specified below (see Table \ref{tab:input}).

$P_i$ are the same as in the SM while the short distance physics is now described by the loop functions
\be
X_K=|X_K|\,e^{i\,\theta_X^K}\,,\qquad Y_K=|Y_K|\,e^{i\,\theta_Y^K}\,,\qquad
Z_K=|Z_K|\,e^{i\,\theta_Z^K}\,,\qquad E_K=|E_K|\,e^{i\,\theta_E^K}\,,
\ee
that are generalizations of the real valued SM loop functions $X_0$, $Y_0$, $Z_0$ and $E_0$ in \eqref{epeSM} to the LHT model. Explicit expressions for $X_K$, $Y_K$ and $Z_K$ have been obtained in \cite{Blanke:2006eb}. The function $E_K$ can be found, in complete analogy to the functions $T_{D'}$ and $T_{E'}$ governing the $B\to X_s\gamma$ decay \cite{BBPTUW}, by changing the argument of the SM $E_0$ function and properly adjusting various overall factors. The result is given in Appendix \ref{sec:appA}. The phases $\beta_i^K$ entering \eqref{FE} are then given by
\be
\beta_i^K = \beta -\beta_s - \theta^K_i\qquad (i = X,Y,Z,E)\,.
\ee

A comment on two approximations made above is in order. The first one concerns the contributions from the T-even sector to the functions $X_K$ and $Y_K$. In the calculation of these functions, the fermion mass on the flavour conserving side of the box diagrams has been set to zero, since in the case of semileptonic rare decays SM leptons are present. On the other hand, in the case of non-leptonic decays, such as $K_L\to\pi\pi$, this mass cannot be generally neglected, as now up-type quarks, in particular the top quark and the heavy $T_+$, contribute. However, it can straightforwardly be shown that including this difference results in the presence of a new operator \cite{BBH}
\be
(\bar s d)_{V-A}(\bar bb)_{V-A}
\label{eq:bop}
\ee
at scales $\mu>m_b$,  which is not contained in \eqref{epeSM} and \eqref{epeth}, \eqref{FE}. 
It is multiplied by the function 
\be
S_t = S_0(x_t) + \bar S_\text{even}\,,
\ee   
where $S_0(x_t)$ denotes the SM contribution and $\bar S_\text{even}$ the heavy $T_+$ contribution.
{Below the scale $\mu=m_b$ the $b$ quark is integrated out, and therefore the
operator in \eqref{eq:bop} contributes to $\epe$ only through mixing under renormalization.} In the case of the SM, this contribution has been shown to be $\ord(1\%)$ and therefore fully negligible \cite{BBH}. As in the LHT model the dominant contribution to $S_t$ comes from the SM part $S_0(x_t)$ \cite{BBPTUW,BPU}, the accuracy of neglecting this contribution remains the same in the LHT model.

The second approximation entering the above formula \eqref{FE} concerns the
T-odd sector and consists in neglecting the mass splittings of mirror quarks
on the flavour conserving side of the box diagrams contributing to the $X_K$
and $Y_K$ functions. We have checked, see also \cite{Blanke:2006eb}, that the
inclusion of these splittings affects the functions $X_K$ and $Y_K$ by at most
10\%. As $P_X$ and $P_Y$ are much smaller than $P_0$ and $|P_Z|$, these functions do not play a dominant role in $\epe$ anyway and we can safely neglect also this effect in view of large non-perturbative uncertainties.

In the LHT model, the first term in (\ref{FE}), which involves
$P_0$ and is dominated by the 
QCD penguin operator $\mathcal{Q}_6$, does not contain any NP contribution. 
On the other hand, the important negative last term involving $P_Z$ and 
related  to the EW penguin operator $\mathcal{Q}_8$ can be strongly enhanced, when
$\theta_Z \neq 0$, $\sin \beta_Z \simeq 1$ and $|Z| > Z_0(x_t)$.
These conditions can indeed be satisfied, as found in \cite{Blanke:2006eb} from a general scan over the
  three generation mirror fermion masses and the six parameters (three angles
  $\theta^d_{12}$, $\theta^d_{13}$, $\theta^d_{23}$ and three phases
  $\delta^d_{12}$, $\delta^d_{13}$, $\delta^d_{23}$) of the mixing matrix $V_{Hd}$.
Thus, in this case, the suppression of $\epe$ through
the enhanced electroweak penguin contribution must be compensated by the
increase of the QCD penguin contribution $P_0$ or by decreasing the magnitude
of the coefficient $P_Z$.
This corresponds to the increase of $R_6$ and the decrease of $R_8$, respectively.

Clearly, as seen in the previous section, the result for $\epe$ is 
very sensitive to the actual values of the coefficients $P_i$. In the LHT model, in addition, there is a strong dependence on the phases $\beta_i^K$.

We conclude this section commenting on the origin of the correlations present in the LHT
model between $\epe$ and rare kaon decays.
They come from the simultaneous dependence of rare $K$ decays and
$\epe$ on the short-distance functions $X_K$, $Y_K$ and $Z_K$.
For instance, the branching ratio for $\klpn$ reads
\be
\label{eq:BrKL}
Br(K_{L}\to\pi^0\nu\bar\nu) =
\kappa_L\tilde r^2 A^4 R_t^2 |X_K|^2 \sin^2{\beta_X^K}\,, 
\ee
where \cite{BGHN06}
\be
\kappa_L=(2.22\pm0.07)\cdot 10^{-10}\,,\qquad\tilde r = \left|\frac{V_{ts}}{V_{cb}}\right|\simeq 0.98 \,,\qquad
 R_t = \frac{| V_{td}^{}V^*_{tb}|}{| V_{cd}^{}V^*_{cb}|}\simeq 1.0\,.
\ee

As, in the LHT model, there are also strong correlations between $X_K$, $Y_K$
 and $Z_K$, in particular between their phases, it is evident that there will
 be a strong correlation between the CP-violating observables $\epe$ and $Br(\klpn)$.

The explicit expressions for $Br(\kpn)$ and $Br(K_L \to \pi^0 \ell^+ \ell^-)$
in terms of $X_K$, $Y_K$ and $Z_K$ are more complicated than the one in
(\ref{eq:BrKL}). They are given in \cite{Blanke:2006eb}, to which we refer for
details, and forecast as well correlations between $\epe$ and these decays.

\newsection{Numerical Analysis in the LHT Model}
\label{sec:numerics}
In our numerical analysis in the LHT model presented below we have
 used for the determination of the CKM parameters, and in particular
 of $\IM(\lambda_t)$, the tree
 level values of $|V_{ub}|$, $\vcb$, $\lambda$ and $\gamma$ given in
 Table~\ref{tab:input}, as the UTfit values obtained within the SM are clearly not valid in the LHT model.
 In obtaining the SM values of rare decay branching ratios in
 Table~\ref{tab:R6R8} below, however, we consistently used the determination of the
 CKM parameters within the SM.
 As a curiosity we remark that with the CKM values of
 Table~\ref{tab:input}, due to an increased value of $\IM(\lambda_t)$
 with respect to the UTfit determination, the SM
 branching ratios are higher than those given in
 Table~\ref{tab:R6R8} and their central 
 values read 
\begin{gather}
Br(\klpn)_\text{SM}^\text{tree}=4.0\cdot 10^{-11}\,,\qquad 
Br(\kpn)_\text{SM}^\text{tree}=9.5\cdot 10^{-11}\,,\\
Br(K_L\to\pi^0e^+e^-)_\text{SM}^\text{tree}=3.8\cdot 10^{-11}\,,\qquad 
Br(K_L\to\pi^0\mu^+\mu^-)_\text{SM}^\text{tree}=1.5\cdot 10^{-11}\,.
\end{gather}
However, such a procedure would not be fully consistent as the CKM values in Table~\ref{tab:input}  deviate
significantly from the SM UTfit values: the reason is the so-called
``$\sin 2 \beta$ problem'' \cite{BBGT}.

\begin{table}[ht]
\renewcommand{\arraystretch}{1}\setlength{\arraycolsep}{1pt}
\center{\begin{tabular}{|l|l|}
\hline
{\small $G_F=1.16637\cdot 10^{-5} \gev^{-2}$} & {\small$\Delta M_K= 3.483(6)\cdot 10^{-15}\gev$} \\
{\small$\mw= 80.425(38)\gev$} & {\small$\Delta M_d=0.508(4)/ \rm{ps}$\hfill\cite{BBpage}} \\\cline{2-2}
{\small$\alpha=1/127.9$} &{\small $\Delta M_s = 17.77(12)/\text{ps}$\hfill\cite{CDFnew,D0}} \\\cline{2-2}
{\small$\sin^2 \theta_W=0.23120(15)$\qquad\hfill\cite{PDG}} & {\small $S_{\psi K_S}=0.675(26)$ \hfill\cite{BBpage}}\\\hline
{\small$|V_{ub}|=0.00409(25)$} &  {\small
  $F_K\sqrt{\hat B_K}= 143(7)\mev$\qquad\hfill\cite{Hashimoto,PDG}}\\\cline{2-2}
{\small $\vcb = 0.0416(7)$\hfill\cite{BBpage}} & {\small$F_{B_d} \sqrt{\hat B_{B_d}}= 214(38)\mev$} \\\cline{1-1}
{\small$\lambda=|V_{us}|=0.2258(14)$}  & {\small$F_{B_s} \sqrt{\hat B_{B_s}}= 262(35)\mev$\;\;\hfill\cite{Hashimoto}} \\\cline{2-2}
 {\small$\gamma=82(20)^\circ$ \hfill\cite{UTFIT}} & {\small$\eta_1=1.32(32)$\hfill\cite{eta1}} \\\hline
{\small$m_{K^0}= 497.65(2)\mev$} & {\small$\eta_3=0.47(5)$\hfill\cite{eta3}}\\\cline{2-2}
{\small$m_{D^0}=  1.8645(4)\gev$} &{\small$\eta_2=0.57(1)$} \\
{\small$m_{B_d}= 5.2794(5)\gev$} & {\small$\eta_B=0.55(1)$\hfill\cite{eta2B}}\\\cline{2-2}
{\small$m_{B_s}= 5.370(2)\gev$} & {\small$\mcb= 1.30(5)\gev$} \\
{\small $|\varepsilon_K|=2.284(14)\cdot 10^{-3}$ \hfill\cite{PDG}} &{\small$\mtb= 161.7(20)\gev$} \\
\hline
\end{tabular}  }
\caption {\textit{Values of the experimental and theoretical
    quantities used as input parameters.}}
\label{tab:input}
\renewcommand{\arraystretch}{1.0}
\end{table}

\begin{figure}[!]
\begin{minipage}{8cm}
\begin{center}
\epsfig{file=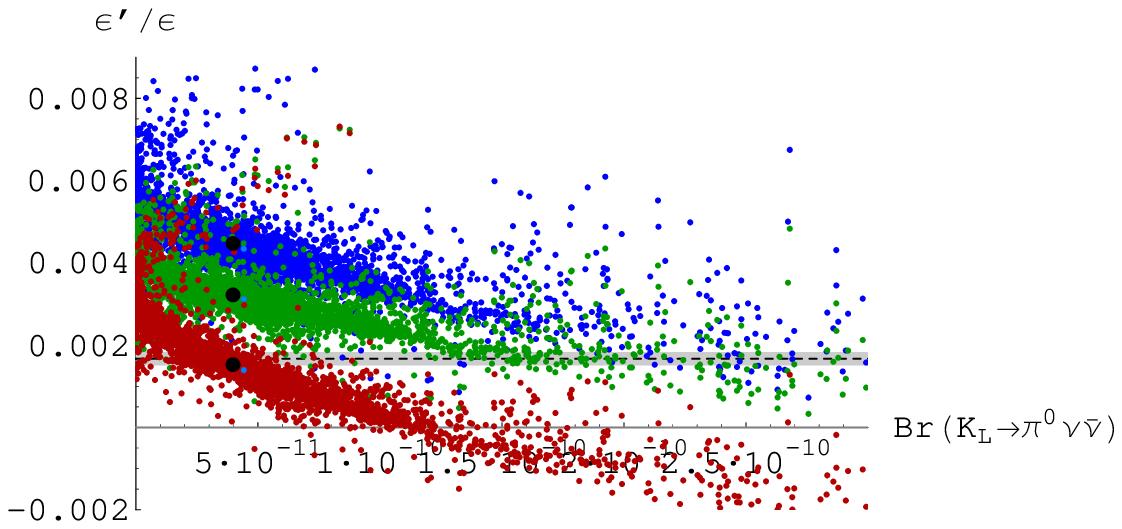,scale=.67}
\end{center}
\end{minipage}
\begin{minipage}{7cm}
\begin{center}
\epsfig{file=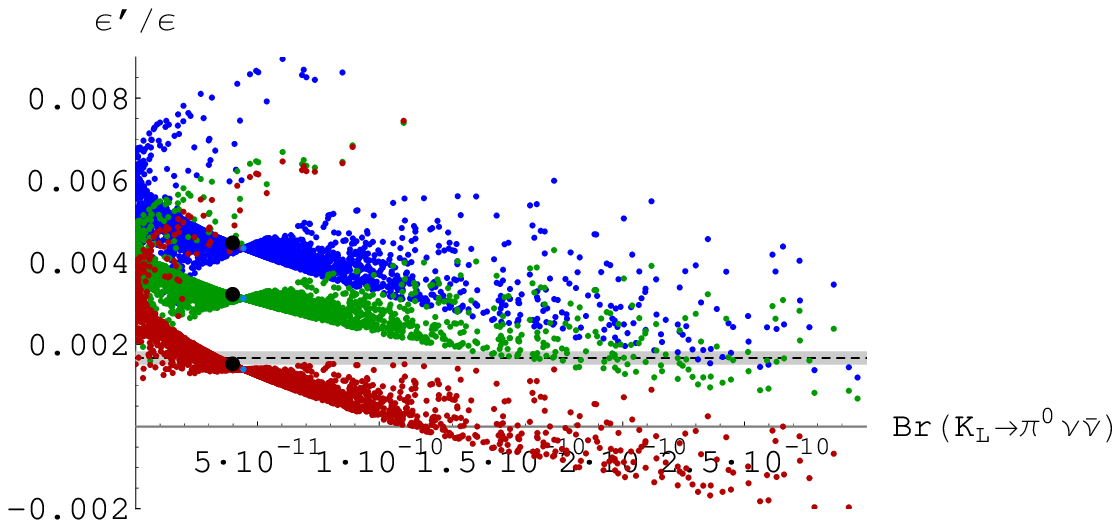,scale=0.67}
\end{center}
\end{minipage}
\caption{\label{fig:eps-KL}\it Left: $\epe$ as a function of $Br(\klpn)$ for
  different values of $(R_6,R_8)=(1.0,1.0)\text{ (red), }(1.5,0.8)\text{
    (green), }(2.0,1.0)\text{ (blue)}$. The shaded area represents the
  experimental result in \eqref{eps} {while the SM predictions are displayed by
    the black points}. Right: Same as before, but with
    two phases ($\delta^d_{12}$ and $\delta^d_{23}$) of the mixing matrix
    $V_{Hd}$  set to zero.
Comparing the left and right plots, it is evident that $\epe$ turns out to be
    quite sensitive to these phases.}
\end{figure}

The  discussion of Sections \ref{sec:SM} and \ref{sec:epsprime} forecasts that in
order to allow large enhancements of the rare decays $\klpn$ and
$K_L\to\pi^0\ell^+\ell^-$, the consistency with the data on $\epe$ requires
$R_6>R_8$. In Fig.~\ref{fig:eps-KL} we show $\epe$ as a function of
$Br(\klpn)$ {in the LHT model} for different values of $(R_6,R_8)$. To this end we have set
$\Lms=340\mev$ and performed a general scan over the parameters of the LHT
model subject to present experimental constraints from $K$ and $B$ physics as
discussed in detail in \cite{Blanke:2006eb,BBPTUW}.
We compare the plot resulting from the general scan with the one obtained
setting to zero two phases, $\delta^d_{12}$ and $\delta^d_{23}$, of the $V_{Hd}$
mixing matrix.\footnote{A detailed analysis of the number of phases in the mixing matrices
in the LHT model has been presented in \cite{SHORT}.}
These two plots are significantly different, signaling that $\epe$ is quite
sensitive to the new phases $\delta^d_{12}$ and $\delta^d_{23}$, {whereas this sensitivity was much weaker in the case of rare decays
 discussed in \cite{Blanke:2006eb}.
This shows that $\epe$ is not only very sensitive
to the values of the hadronic matrix elements but also to the new parameters of
 a given model. This fact could be used in the future to efficiently
 exclude some portions of the parameter space provided the hadronic matrix
 elements will be brought under control.}

We observe that for $(R_6,R_8)=(1.0,1.0)$ (red points), large enhancements of
$Br(\klpn)$ over the SM value imply a strong suppression of $\epe$ relative to
the data, and consequently in this case large enhancements of $Br(\klpn)$
found in the LHT model in \cite{Blanke:2006eb} are unlikely. The same applies to
$Br(K_L\to\pi^0\ell^+\ell^-)$. On the other hand, for $(R_6,R_8)=(1.5,0.8)$
(green points) and $(R_6,R_8)=(2.0,1.0)$ (blue points) the experimental data for $\epe$ imply in the LHT model a significant enhancement of $Br(\klpn)$ with respect to the SM.

As $\klpn$ and $K_L\to\pi^0\ell^+\ell^-$ are very strongly correlated with each other \cite{Blanke:2006eb}, also $Br(K_L\to\pi^0\ell^+\ell^-)$ are predicted to be enhanced for $(R_6,R_8)=(1.5,0.8)$ and $(R_6,R_8)=(2.0,1.0)$. We summarize in Table~\ref{tab:R6R8} the three choices for $(R_6,R_8)$ and the corresponding values of rare decay branching ratios that are compatible with the data for $\epe$.

\begin{table}[ht]
{\renewcommand{\arraystretch}{1.1}\setlength{\arraycolsep}{2pt}
\begin{center}
\begin{tabular}{|c || c|c|c|c|}
\hline
 & SM & $(1.0,1.0)$ & $(1.5,0.8)$ & $(2.0,1.0)$ \\ \hline
$Br(\klpn)\cdot 10^{11}$ & $2.7\pm0.4$& 0.007\dots 9.5 & 0.5\dots 43 & 8.4\dots 42 \\
$Br(\kpn)\cdot 10^{10}$ & $0.84\pm0.10$& 0.09\dots 5.7 & 0.6\dots
 2.3 & 1.0\dots 1.8 \\
$Br(K_L\to\pi^0e^+e^-)\cdot 10^{11}$ & $3.54^{+0.62}_{-0.49}$& 2.7\dots4.7 & 2.9\dots8.8 & 4.2\dots8.6 \\
 $Br(K_L\to\pi^0\mu^+\mu^-)\cdot 10^{11}$ & $1.41^{+0.28}_{-0.26}$& 1.2\dots1.8 & 1.2\dots3.9 & 1.8\dots3.8 \\\hline
\end{tabular}
\end{center}}
\caption{\label{tab:R6R8}\it  Choices for $(R_6,R_8)$ and the corresponding
 values of rare decay branching ratios that are compatible with the data for
 $\epe$. The SM predictions \cite{cern2} are also shown. {For
 $Br(K_L\to\pi^0\ell^+\ell^-)$ we consider for simplicity only the case of
 constructive interference between direct and indirect CP-violation.}}
\end{table}

\begin{figure}[!]
\begin{center}
\epsfig{file=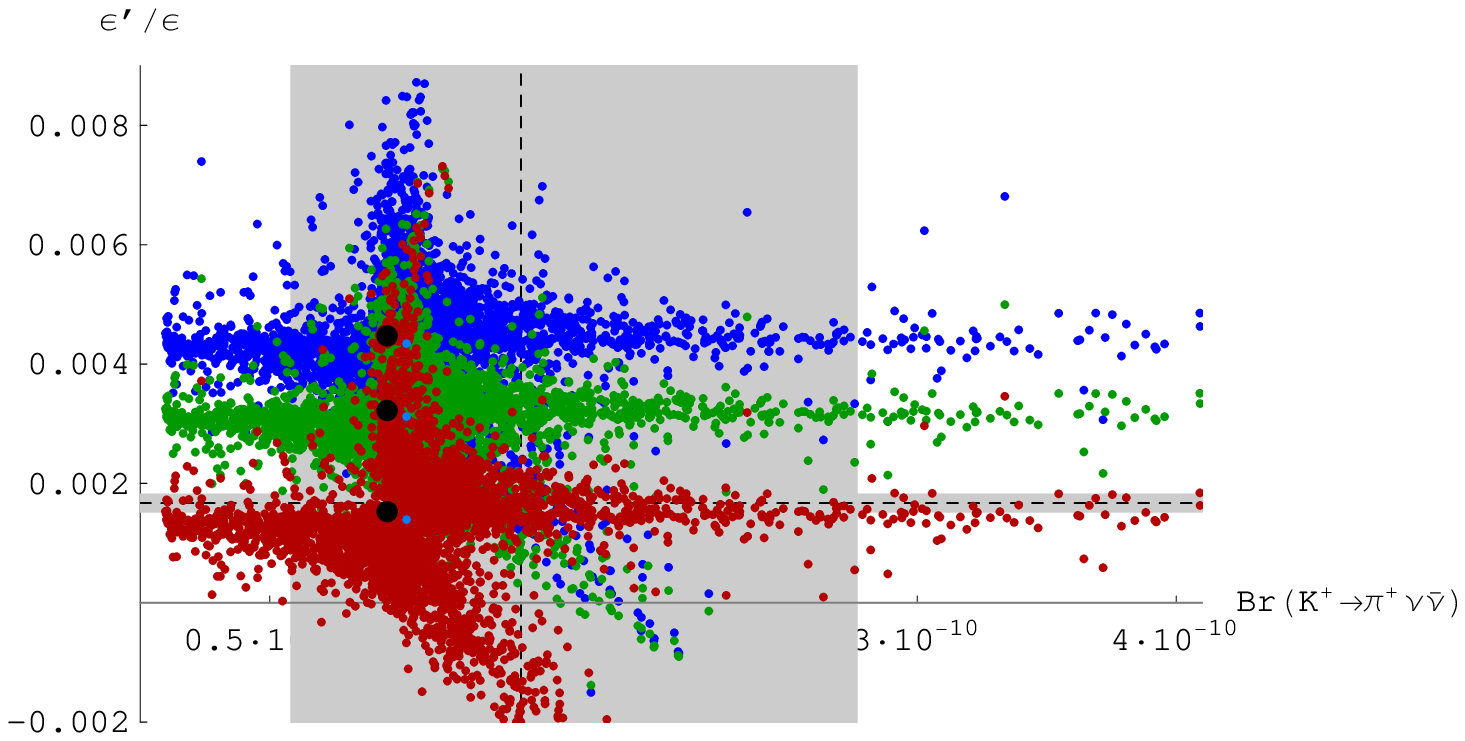,scale=0.8}
\end{center}
\caption{\label{fig:eps-K+}\it Correlation between $\epe$ and $\kpn$ for
  different values of $(R_6,R_8)=(1.0,1.0)\text{ (red), }(1.5,0.8)\text{
    (green), }(2.0,1.0)\text{ (blue)}$. The shaded areas represent the
  experimental results {while the SM predictions are displayed by
    the black points.}}
\end{figure}

\begin{figure}[!]
\begin{minipage}{7.7cm}
\begin{center}
\epsfig{file=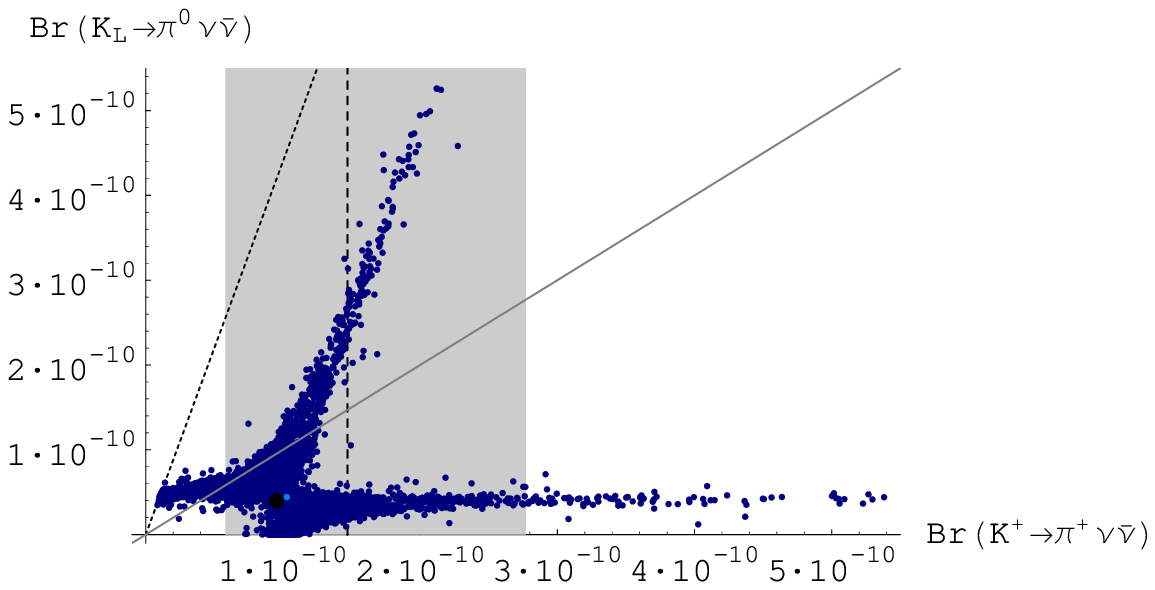,scale=.67}
\end{center}
\end{minipage}
\begin{minipage}{7.2cm}
\begin{center}
\epsfig{file=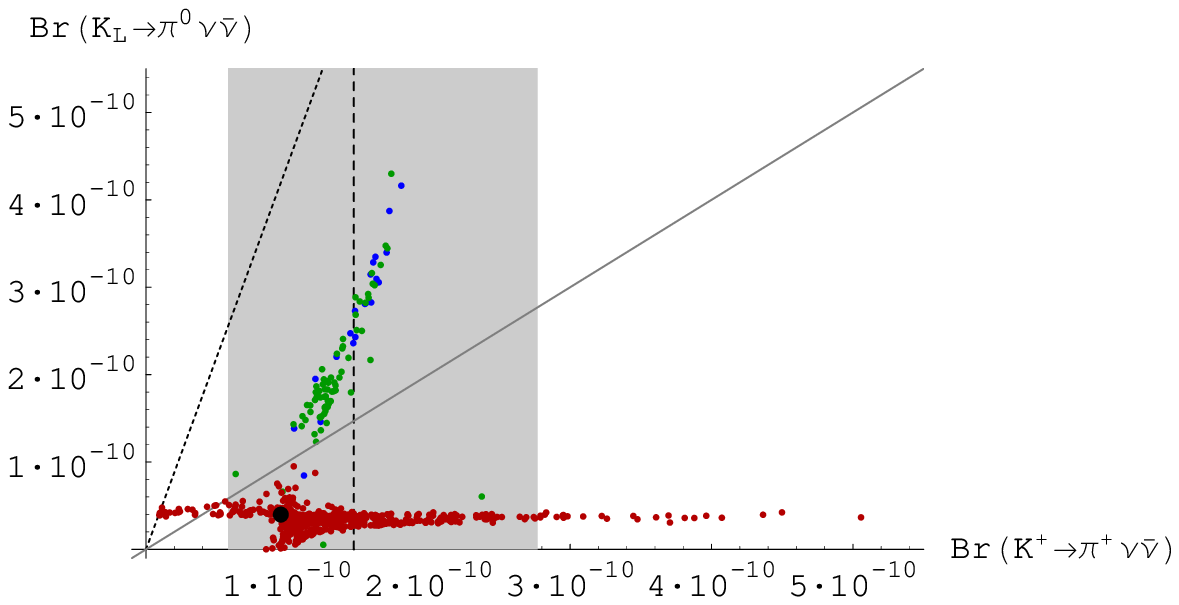,scale=0.67}
\end{center}
\end{minipage}
\caption{\label{fig:KL-K+}\it Left: Correlation between $\kpn$ and $\klpn$
  without imposing the $\epe$-constraint \cite{Blanke:2006eb}. The shaded area
  represents the experimental result for $Br(\kpn)$ \cite{expK+} {while the SM predictions are displayed by
    the black points.} The
  Grossman-Nir bound \cite{Grossman:1997sk} is displayed by the dotted line, while the solid line separates the two areas where $Br(\klpn)$ is larger or smaller than $Br(\kpn)$. Right: Same as before, but after imposing the constraint from $\epe$ with different values of $(R_6,R_8)=(1.0,1.0)\text{ (red), }(1.5,0.8)\text{ (green), }(2.0,1.0)\text{ (blue)}$.}
\end{figure}

In Fig.~\ref{fig:eps-K+} we show the correlation between $\epe$ and $\kpn$ that is significantly weaker than in the case of $\epe$ and $\klpn$. 
In particular, we find that in the case $(R_6,R_8)=(1.0,1.0)$, in which
$Br(\klpn)$ and $Br(K_L\to\pi^0\ell^+\ell^-)$ are required to be SM-like,
$Br(\kpn)$ can be largely enhanced relative to its SM value. A different
behaviour is observed for the two other choices of $(R_6,R_8)$ considered by
us. Here only enhancements of $Br(\kpn)$ by at most a factor $3$ are allowed.

In order to understand better the pattern of enhancements of $Br(\klpn)$ and $Br(\kpn)$, we show in Fig.~\ref{fig:KL-K+} the correlation between $Br(\klpn)$ and $Br(\kpn)$ in the LHT model without the $\epe$ constraint as obtained in \cite{Blanke:2006eb}, and after the constraint from $\epe$  for different choices for $(R_6,R_8)$ has been taken into account. We observe that setting $(R_6,R_8)=(1.0,1.0)$ basically selects the horizontal branch on which $Br(\klpn)$ is SM-like but $Br(\kpn)$ can be strongly enhanced. The other two choices for $(R_6,R_8)$ select the second branch on which $Br(\klpn)$ can be strongly enhanced but $Br(\kpn)<2.3\cdot10^{-10}$.

\newsection{Conclusions}
\label{sec:conclusions}

In this paper we have calculated $\epe$ for different values of the hadronic
parameters $(R_6,R_8)$ in the LHT model and investigated the implications for
rare decay branching ratios when taking the experimental data for $\epe$ into
account. The main results of our paper are given in
Figs.~\ref{fig:eps-KL}--\ref{fig:KL-K+} and in Table \ref{tab:R6R8} and
can be summarized as follows:
\bi
\item
For the values of hadronic parameters $(R_6,R_8)\simeq(1.0,1.0)$, for which
$(\epe)_\text{SM}$ agrees with the data, large enhancements of $Br(\klpn)$ and
$Br(K_L\to\pi^0\ell^+\ell^-)$ relative to the SM are very unlikely.
\item
On the other hand, for the values of hadronic parameters
  $(R_6,R_8) = (1.5,0.8)$ and $(2.0,1.0)$ chosen by us, the large NP contributions that are
  required to fit the experimental value for $\epe$ result in large
  enhancements of $Br(\klpn)$ and $Br(K_L\to\pi^0\ell^+\ell^-)$
  relative to the SM.
\item
The correlation between $\epe$ and $\kpn$ is much weaker and large departures of $Br(\kpn)$ from the SM values are possible even for  $(R_6,R_8)\simeq(1.0,1.0)$, however, more modest enhancements are possible for the other choices of hadronic parameters, as seen in Figs.~\ref{fig:eps-K+} and \ref{fig:KL-K+}.
\ei

The main message of our paper is clear: without significant progress in the evaluation of $R_6$ and $R_8$ and other less important hadronic parameters entering $\epe$, the role of the data in \eqref{eps} in constraining physics beyond the SM will remain limited.

\subsection*{Acknowledgments}
We thank S. Bethke for enlighting information on the status of $\alpha_s(M_Z)$.
This research was partially supported by the Cluster of Excellence `Origin and Structure of the Universe' and by the German Bundesministerium f\"ur Bildung und Forschung under contract 05HT6WOA.

\begin{appendix}

\newsection{Explicit Formulae for the Function $\bm{E_K}$}\label{sec:appA}

In this appendix we give the explicit expression for the function $E_K$
entering the calculation of $\epe$ in the LHT model. The functions $X_K$,
$Y_K$ and $Z_K$ have been calculated already in \cite{Blanke:2006eb} in the context
of rare $K$ and $B$ decays and can be found in that paper. The variables are defined as follows:
\begin{gather}
x_t=\frac{m_t^2}{M_{W_L}^2}\,,\qquad x_T =\frac{m_{T_+}^2}{M_{W_L}^2}\,,\\
z_i = \frac{m_{Hi}^2}{M_{W_H}^2}\,,\qquad z'_i = a\,z_i\quad \text{with }a=\frac{5}{\tan^2\theta_W} \qquad (i=1,2,3)\,,\\
\lambda_t = V_{ts}^* V_{td}^{}\,,\qquad \xi_i^{(K)} = V_{Hd}^{*is}V_{Hd}^{id} \qquad (i=1,2,3)\,,
\end{gather}
and $x_L$ describes the mixing in the T-even top sector.

\bea
E_0(x_t) &=&  -\frac{2}{3}\log{x_t}+\frac{x_t^2(15-16x_t+4x_t^2)}{6(1-x_t)^4}\log{x_t} + \frac{x_t(18-11x_t-x_t^2)}{12(1-x_t)^3}\,,\\
E_K &=& E_0(x_t) + \bar E_\text{even} + \frac{1}{\lambda_t}\bar E^\text{odd}_K\,,\\
\bar E_\text{even} &=& x_L^2 \frac{v^2}{f^2}\left[E_0(x_T)- E_0(x_t)\right]\,,\\
\bar E^\text{odd}_K &=& \frac{1}{4}\frac{v^2}{f^2}\sum_i \xi_i^{(K)}\left[\frac{3}{2} E_0(z_i) + \frac{1}{10}E_0(z'_i) \right]\,.
\eea

\end{appendix}

\end{document}